\documentclass[twocolumn]{aastex63}

\usepackage{amsmath}

\shorttitle{ATLAS's First TDE}
\shortauthors{Hinkle et al.}

\graphicspath{{./}{figures/}}

\begin{document}

\title{SCAT Uncovers ATLAS's First Tidal Disruption Event ATLAS18mlw: A Faint and Fast TDE in a Quiescent Balmer Strong Galaxy}

\correspondingauthor{Jason T. Hinkle}
\email{jhinkle6@hawaii.edu}

\author[0000-0001-9668-2920]{Jason T. Hinkle}
\affiliation{Institute for Astronomy, University of Hawai`i, 2680 Woodlawn Drive, Honolulu, HI 96822, USA}

\author[0000-0002-2471-8442]{Michael A. Tucker}
\altaffiliation{DOE CSGF Fellow}
\affiliation{Institute for Astronomy, University of Hawai`i, 2680 Woodlawn Drive, Honolulu, HI 96822, USA}

\author[0000-0003-4631-1149]{Benjamin. J. Shappee}
\affiliation{Institute for Astronomy, University of Hawai`i, 2680 Woodlawn Drive, Honolulu, HI 96822, USA}

\author[0000-0001-9206-3460]{Thomas W.-S. Holoien}
\altaffiliation{NHFP Einstein Fellow}
\affiliation{The Observatories of the Carnegie Institution for Science, 813 Santa Barbara Street, Pasadena, CA 91101, USA}

\author[0000-0001-5661-7155]{Patrick J. Vallely}
\affiliation{Department of Astronomy, The Ohio State University, 140 West 18th Avenue, Columbus, OH 43210, USA}

\author{Thomas de Jaeger}
\affiliation{Institute for Astronomy, University of Hawai`i, 2680 Woodlawn Drive, Honolulu, HI 96822, USA}

\author[0000-0002-4449-9152]{Katie~Auchettl}
\affiliation{School of Physics, The University of Melbourne, Parkville, VIC 3010, Australia}
\affiliation{ARC Centre of Excellence for All Sky Astrophysics in 3 Dimensions (ASTRO 3D)}
\affiliation{Department of Astronomy and Astrophysics, University of California, Santa Cruz, CA 95064, USA}

\author{Greg Aldering}
\affiliation{Lawrence Berkeley National Laboratory, 1 Cyclotron Rd., Berkeley, CA, 94720, USA}

\author{Chris Ashall}
\affiliation{Institute for Astronomy, University of Hawai`i, 2680 Woodlawn Drive, Honolulu, HI 96822, USA}

\author[0000-0002-2164-859X]{Dhvanil D. Desai}
\affiliation{Institute for Astronomy, University of Hawai`i, 2680 Woodlawn Drive, Honolulu, HI 96822, USA}

\author{Aaron Do}
\affiliation{Institute for Astronomy, University of Hawai`i, 2680 Woodlawn Drive, Honolulu, HI 96822, USA}

\author[0000-0003-3490-3243]{Anna V. Payne}
\altaffiliation{NASA Fellowship Activity Fellow}
\affiliation{Institute for Astronomy, University of Hawai`i, 2680 Woodlawn Drive, Honolulu, HI 96822, USA}

\author{John L. Tonry}
\affiliation{Institute for Astronomy, University of Hawai`i, 2680 Woodlawn Drive, Honolulu, HI 96822, USA}

\begin{abstract}
\noindent We present the discovery that ATLAS18mlw was a tidal disruption event (TDE) in the galaxy WISEA J073544.83+663717.3, at a luminosity distance of 334 Mpc. Initially discovered by the Asteroid Terrestrial Impact Last Alert System (ATLAS) on 2018 March 17.3, the TDE nature of the transient was uncovered only recently with the re-reduction of a SuperNova Integral Field Spectrograph (SNIFS) spectrum. This spectrum, taken by the Spectral Classification of Astronomical Transients (SCAT) survey, shows a strong blue continuum and a broad H$\alpha$ emission line. Here we present roughly six years of optical survey photometry beginning before the TDE to constrain AGN activity, optical spectroscopy of the transient, and a detailed study of the host galaxy properties through analysis of archival photometry and a host spectrum. ATLAS18mlw was detected in ground-based light curves for roughly two months. From a blackbody fit to the transient spectrum and bolometric correction of the optical light curve, we conclude that ATLAS18mlw is best explained by a low-luminosity TDE with a peak luminosity of log(L [erg s$^{-1}$]) = $43.5 \pm 0.2$. The TDE classification is further supported by the quiescent Balmer strong nature of the host galaxy. We also calculated the TDE decline rate from the bolometric light curve and find $\Delta L_{40} = -0.7 \pm 0.2$ dex, making ATLAS18mlw a member of the growing class of ``faint and fast'' TDEs with low peak luminosities and fast decline rates.
\end{abstract}


\section{Introduction} \label{sec:intro}

When a star passes within the tidal radius of a supermassive black hole (SMBH), the self-gravity of the star is overwhelmed by tidal forces and the star is disrupted, resulting in a tidal disruption event \citep[TDE; ][]{rees88, evans89, phinney89}. A TDE results in a luminous flare as the disrupted stellar material falls back onto the SMBH \citep[e.g.,][]{ulmer99}. To date, $\sim60$ of these flares have been observed \citep{gezari21}, with a majority discovered in the optical. This is in large part due to the recent expansion of transient surveys such as the All-Sky Automated Survey for Supernovae \citep[ASAS-SN;][]{shappee14, kochanek17}, the Asteroid Terrestrial Impact Last Alert System \citep[ATLAS;][]{tonry18}, Gaia Alerts \citep{wyrzykowski12}, the Panoramic Survey Telescope and Rapid Response System \citep[Pan-STARRS;][]{chambers16}, and the Zwicky Transient Facility \citep[ZTF;][]{bellm19}.

However, it is not only optical sky surveys that play an important role in discovering TDEs. Spectroscopic classification surveys are crucial in obtaining rapid spectra to not only classify transients but to also identify interesting sources. Alongside the expansion of transient surveys has been the growth of such spectroscopic surveys. In the past two years, roughly three-quarters of all spectroscopically-confirmed transients have been classified by the Zwicky Transient Facility classification efforts \citep[e.g.,][]{fremling20, fremling21}, the Public ESO Spectroscopic Survey for Transient Objects \citep[PESSTO;][]{smartt15} and successors \citep[e.g.,][]{barbarino19}, and the Spectral Classification of Astronomical Transients (SCAT) survey \citep{tucker22}, in that order.

Spectroscopic classification of TDEs has become easier as our understanding of the typical observed emission from TDEs becomes more complete. The emission of TDEs is hot, typically exhibiting a blue optical continuum, with a UV excess. The UV/optical spectral energy distribution (SED) of TDEs is often well-fit by a blackbody with a temperature on the order of T $= 20000 - 50000$ K \citep[e.g.,][]{holoien14b, holoien16a, blagorodnova17, hinkle21b}. Roughly half of optically-selected TDEs show X-ray emission, commonly in the form of a blackbody with kT $\sim 50 $ eV, or a soft power-law with a photon index of $\gtrsim 2.5$ \citep{brown17a, auchettl17, auchettl18}. The characteristic lifetime of a TDE in the optical bands is on the order of a few months \citep[e.g.,][]{hinkle20a, vanvelzen20b} although UV emission can exist above host galaxy levels for several years after disruption \citep[e.g.,][]{vanvelzen19b}. 

TDEs represent an excellent opportunity to study otherwise non-active SMBHs. For example, TDE emission is likely sensitive to black hole spin and mass \citep[e.g.][]{ulmer99, graham01, mockler19, gafton19}. It has been shown that the SMBH masses obtained through fits to TDE light curves are consistent with those derived through other means \citep{mockler19}. As the large majority of SMBHs are not active \citep[$\sim 90$\%; ][]{ho08, lacerda20}, this agreement suggests that TDEs may provide an important method of probing quiescent SMBHs at large distances.

Unfortunately, TDEs are rare, with observational rates estimated to be between $10^{-4}$ and $10^{-5} \text{ yr}^{-1}$ per galaxy \citep[e.g.][]{vanvelzen14, holoien16a, vanvelzen18, auchettl18}. Interestingly, TDEs seem to prefer particular host galaxies where a recent burst of star formation has occurred \citep[e.g.,][]{stone16b, french16} and moderately massive stars may be preferred \citep{mockler22}. These include quiescent Balmer strong (QBS) and post-starburst (PSB) galaxies, both likely post-merger systems \citep{french21}. TDEs occur at rates enhanced by factors of roughly 20 and 40 times the typical values for QBS and PSB galaxies respectively \citep[e.g.][]{arcavi14, french16, law-smith17, graur18, french20b}. Furthermore, \citep{arcavi22} find that the relative contamination of Type Ia SNe to TDEs is decreased by a factor of $\sim 2.7$ in QBS or PSB galaxies as compared to quiescent galaxies.

Due in part to the preference for TDEs to occur in unique galaxies and their slightly increased rates around lower mass SMBHs \citep{wang04, stone16a, vanvelzen18}, the host galaxies of TDEs have received special attention \citep{law-smith17, french20, french20b}. To fulfill the promise of TDEs as probes of SMBHs, studies of the history of accretion, mergers, and star-formation in the TDE host galaxies are needed \citep[e.g.,][]{prieto16}.

Here we present the observations of the first TDE discovered by ATLAS, ATLAS18mlw. The paper is organized in the following manner. In Section~\ref{sec:data} we detail observations of the TDE candidate and host galaxy. In Section \ref{sec:host} we discuss the properties of the host galaxy of ATLAS18mlw. In Section \ref{sec:analysis} we analyze the data on this source, and in Section \ref{sec:discusssion} we discuss our results. Throughout this paper, we have used a cosmology with $H_0$ = 69.6 km s$^{-1}$ Mpc$^{-1}$, $\Omega_{M} = 0.29$, and $\Omega_{\Lambda} = 0.71$ \citep{wright06, bennett14}.

\section{Data}
\label{sec:data}

\subsection{Archival Photometry}

We obtained archival photometry of the host galaxy WISEA J073544.83+663717.3 including $ugriz$ magnitudes from the Sloan Digital Sky Survey (SDSS) Data Release 15 \citep{aguado19} and $W1$ and $W2$ magnitudes from the Wide-field Infrared Survey Explorer \citep[WISE;][]{wright10} AllWISE catalog. While the host galaxy was not listed in any Galaxy Evolution Explorer \citep[GALEX;][]{martin05} catalog, we measured a 6\farcs{0} radius aperture $FUV$ and $NUV$ limit and magnitude from GALEX images using gPhoton \citep{million16}. These ultraviolet, optical, and infrared host-galaxy magnitudes and estimated uncertainties are presented in Table~\ref{tab:arch_phot}.

\begin{table}
\centering
 \caption{Archival Host-Galaxy Photometry}
 \label{tab:arch_phot}
 \begin{tabular}{ccc}
  \hline
  Filter & Magnitude & Magnitude Uncertainty\\
  \hline
  $FUV$ & $>$23.56 & 3$\sigma$ \\
  $NUV$ & 23.66 & 0.34 \\
  $u$ & 21.76 & 0.23 \\
  $g$ & 20.08 &	0.03 \\
  $r$ & 19.34 & 0.02 \\
  $i$ & 18.96 & 0.02 \\
  $z$ & 18.79 & 0.06 \\
  $W1$ & 19.04 & 0.06 \\
  $W2$ & 19.46 & 0.16 \\
  \hline
 \end{tabular}\\
\begin{flushleft}Archival magnitudes of the host galaxy WISEA J073544.83+663717.3. The $FUV$ and $NUV$ magnitudes were measured from GALEX images, the $ugriz$ magnitudes are taken from the SDSS catalog and the $W1$ and $W2$ magnitudes are taken from the AllWISE catalog. All magnitudes are presented in the AB system. \end{flushleft}
\end{table}

\subsection{Transient Imaging}

\begin{figure*}
\centering
 \includegraphics[width=1.0\textwidth]{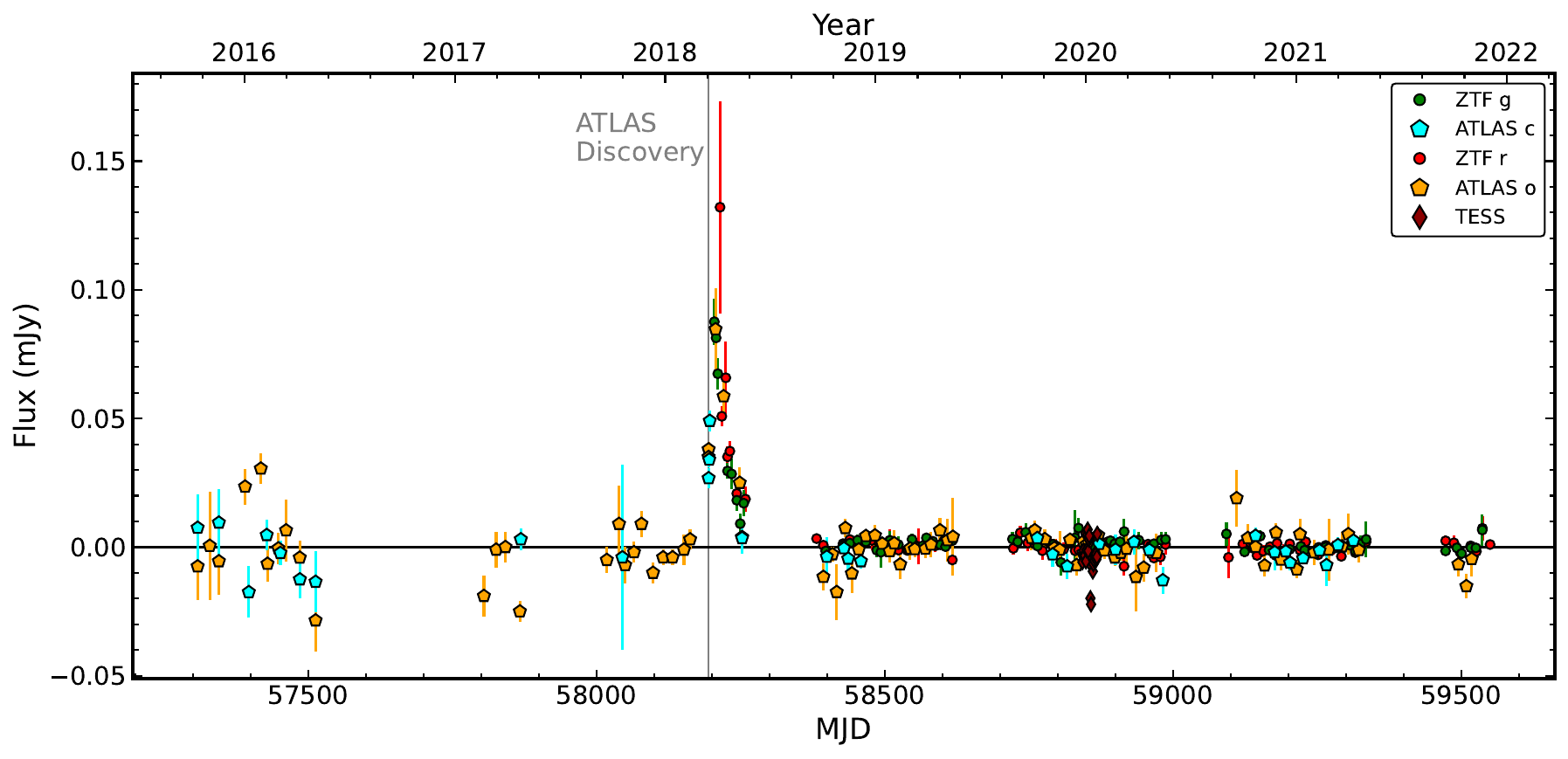}\hfill
 \caption{Host-subtracted and Galactic extinction-corrected light curve of the host galaxy WISEA J073544.83+663717.3 from ATLAS, ZTF, and TESS spanning approximately six years. The ATLAS18mlw flare is clearly visible during early 2018. The green and red circles represent ZTF $g$- and $r$-band photometry stacked in 1-day bins during the flare and 10-day bins outside of the flare. The cyan and orange circles represent ATLAS $c$- and $o$-band photometry stacked in 1-day bins during the flare and 10-day bins outside of the flare. The dark red circles represent TESS photometry stacked in 1-day bins. The vertical gray line marks the epoch of ATLAS discovery.}
 \label{fig:long_lc}
\end{figure*}

The transient ATLAS18mlw, $(\alpha,\delta) =$ (07:35:44.777, $+$66:37:16.43), was discovered on 2018 March 17.3 by the ATLAS survey. The discovery was announced publicly on the Transient Name Server (TNS) and given the name AT2018ahl\footnote{\url{https://www.wis-tns.org/object/2018ahl}} \citep{tonry18ahl}. To give appropriate credit to the discovering survey, we will refer to the transient by its survey name ATLAS18mlw throughout the remainder of this manuscript.

Soon after the ATLAS discovery, we obtained a $V$-band acquisition image with the Supernova Integral Field Spectrograph \citep[SNIFS;][]{lantz04} on the University of Hawai`i 88-in telescope (UH88). From this, we are able to compare the location of the transient to the location of the host galaxy to determine if the source is spatially consistent with the nucleus. We first used astrometry.net \citep{barron08,lang10} to compute astrometry for both the SNIFS acquistion image and a Pan-STARRS $r$-band image \citep{chambers16}. We then measured the centroid of the host galaxy from the Pan-STARRS image and the centroid of the TDE in the SNIFS image with the IRAF task \texttt{imcentroid}. We find a transient position of $(\alpha,\delta)=($07:35:44.793$,+$66:37:16.53$)$ and a host galaxy position of $(\alpha,\delta)=($07:35:44.769$,+$66:37:16.34$)$. Combined, these give an offset of 0\farcs{4}$\pm$0\farcs{5}, where the uncertainty is dominated by the uncertainties in the astrometric solutions for the SNIFS and Pan-STARRS images. At the host distance, this offset corresponds to a physical distance of $600\pm700$~pc, consistent with ATLAS18mlw being a nuclear transient.

We then obtained the ATLAS light curves of ATLAS18mlw from the ATLAS forced point-spread function (PSF) photometry service. ATLAS uses two 0.5-m telescopes located on Haleakal\=a and at the Mauna Loa Observatory. During typical operation, the ATLAS telescopes obtain four 30 s exposures covering roughly a quarter of the sky per night \citep{smith20}. ATLAS obtains images in two broad-band filters, the `cyan' ($c$) filter between 420--650 nm and the `orange' ($o$) filter from 560--820 nm \citep{tonry18}. We combine intra-night epochs using a weighted mean to obtain deeper limits and more robust detections.

In addition to ATLAS photometry, we obtained ZTF photometry in the $g$- and $r$-bands from their forced PSF photometry service. The ZTF survey makes use of the Samuel Oschin 48-in telescope at the Palomar Observatory. In normal operation, ZTF images down to a depth of $\sim 20.5$ $r$-band mag in a 30 s exposure \citep{bellm19}. Similar to the ATLAS data, we combined the intra-night photometric observations using a weighted average to get a single flux measurement.

We also obtained a light curve of the host galaxy from the Transiting Exoplanet Survey Satellite \citep[TESS;][]{ricker15}. TESS observed the location of ATLAS18mlw during Sector 20, at a phase of approximately 650 days after discovery. The TESS full frame images (FFIs) were reduced using the ISIS package \citep{alard98, alard00}. We performed image subtraction following the procedures of \citep[][]{vallely19, vallely21} to compute our differential light curves. The reference image was constructed from the first 100 FFIs of good quality. The measured count rates were converted into fluxes using the instrumental zero point of 20.44 electrons per second \citep{vanderspek18}. To quantify the host variability, we binned the TESS data in 12 hour bins and computed the root mean squared (RMS) scatter, finding a value of 8.1 $\mu$Jy. From this RMS scatter and assuming a variability fraction of 5\%, we find a limit on a possible AGN luminosity in the TESS band of $\lesssim 9\times10^{43}$ erg s$^{-1}$.

Finally, we obtained ASAS-SN photometry in the $g$- and $V$-bands from the newly-public subtracted light curve server\footnote{\url{https://asas-sn.osu.edu/}} \citep{kochanek17}. Over the full baseline of the ASAS-SN survey, the host galaxy of ATLAS18mlw was imaged 285 times in $V$ and 349 in $g$ with no images showing significant detections. The median 5$\sigma$ upper limits are 17.6 and 18.0 mag for the $V$- and $g$-bands respectively.

Figure \ref{fig:long_lc} shows the ATLAS, ZTF, and TESS data for ATLAS18mlw both during the flare and in periods of quiescence. Here the ATLAS and ZTF data have been stacked in 1-day bins during the flare and 10-day bins outside of the flare and the TESS data has been stacked in 1 day bins. Besides the ATLAS18mlw flare itself, it is apparent that no other large flare exists in the roughly six years of coverage. This suggests that the host galaxy did not exhibit AGN-like activity prior to the detection of ATLAS18mlw.

\begin{figure*}
\centering
 \includegraphics[width=0.95\textwidth]{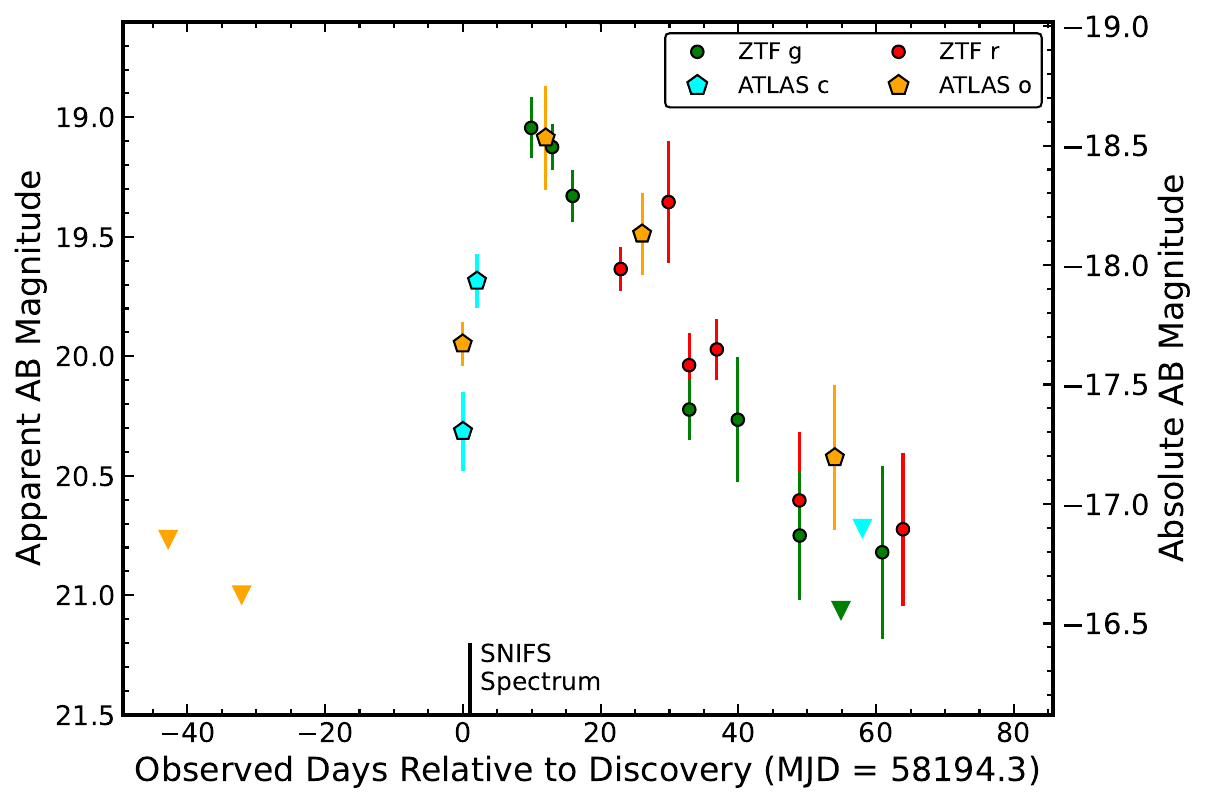}\hfill
 \caption{Host-subtracted and Galactic extinction-corrected optical light curves of ATLAS18mlw. Shown are ATLAS ($co$) and ZTF ($gr$) photometry stacked in 1-day bins. Downward-facing triangles indicate 3$\sigma$ upper limits in filters of the same respective color. The thick black bar on the bottom axis marks the epoch of the initial SNIFS spectrum. All data are shown in the AB magnitude system.}
 \label{fig:opt_lc}
\end{figure*}

The subset of the light curve during the ATLAS18mlw flare is shown in Figure \ref{fig:opt_lc}, showing that ATLAS18mlw rose to peak within $\sim 10$ days after discovery, although there are no non-detections near the discovery epoch to help constrain the time of first light further. After peak, the transient fades by two mags within sixty days and is not detected afterwards. This is faster than most TDEs, although there is no simultaneous UV coverage to test the longevity of the UV excess commonly seen in TDEs \citep[e.g.,][]{holoien16a, vanvelzen20b, gezari21}. 

\subsection{Spectroscopy}

A classification spectrum was obtained 1.05 days after discovery by the SCAT survey \citep{tucker18_scat, tucker22} using SNIFS on UH88. SNIFS is comprised of a blue (B; 320 -- 560 nm) and red (R; 520 nm -- 1000 nm) channel. The original classification spectrum was reduced with a summit-based quick-look pipeline which is a simplified version of the full SNfactory pipeline \citep{aldering06}. The quick-look pipeline estimates the extraction location by averaging over the wavelength axis in each channel and then does aperture photometry.

The quick-look pipeline is reliable when the wavelength dependence of the trace position is less than the seeing. This is true for targets observed near zenith, but atmospheric dispersion shifts the object's location on the chip as the airmass increases. ATLAS18mlw was observed at an airmass of $\approx1.62$ which introduced a wavelength-dependence to the relative contributions from the TDE and the host-galaxy to the final spectrum, masking the transient's signal and hindering classification.

The development of an upgraded reduction pipeline at the University of Hawai`i in June 2021 incorporates wavelength-dependent tracing across each channel. Reanalysis of the classification spectrum revealed a strong blue continuum with narrow host lines and a broad H$\alpha$ line at a redshift of z = 0.073.  A full description of the updated reduction pipeline is presented in \citet{tucker22}. At this redshift, the peak absolute magnitude is $-18.6$ mag. We see none of the broad absorption lines commonly seen in various classes of Type I supernovae \citep{filippenko97}. Additionally, the luminous peak absolute magnitude and lack of a plateau phase in the redder bands likely rule out Type II supernovae \citep[e.g.,][]{galbany16}. Combined, the spectral features and peak absolute magnitude of ATLAS18mlw are consistent with a tidal disruption event \citep[e.g.,][]{vanvelzen20b, gezari21}.

\begin{figure*}
\centering
 \includegraphics[width=0.95\textwidth]{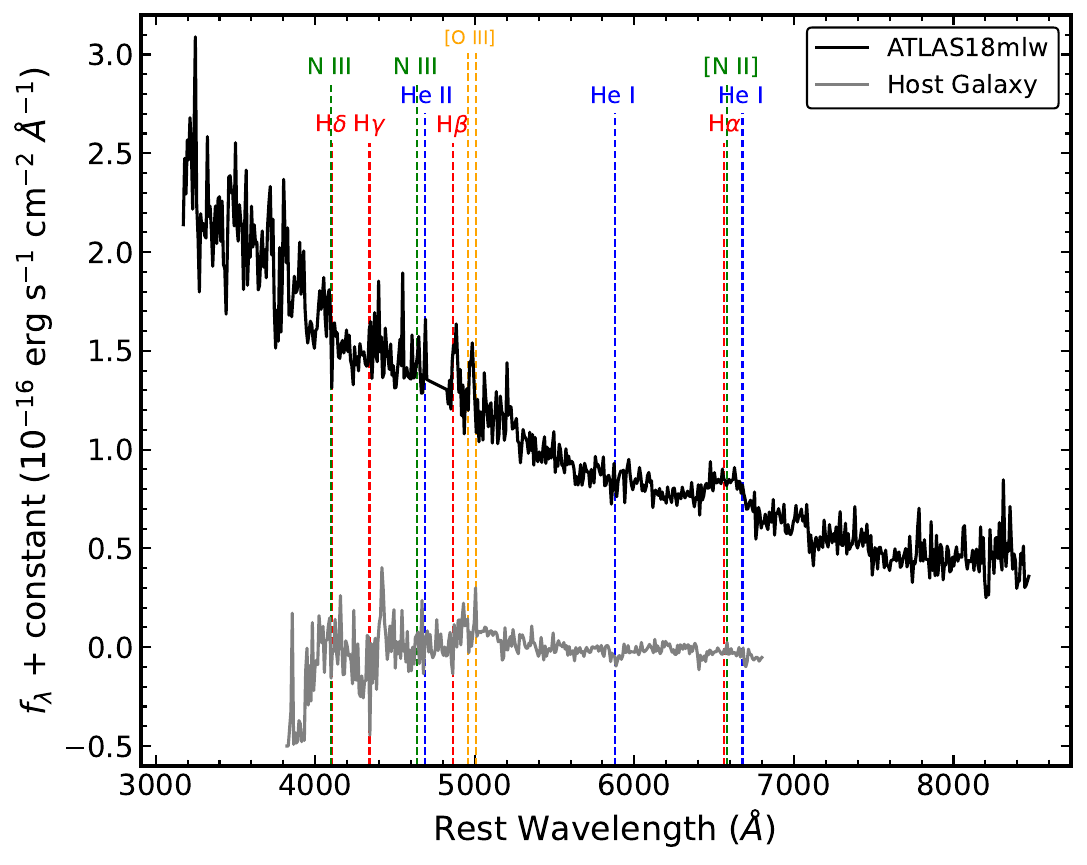}\hfill
 \caption{Optical spectra of ATLAS18mlw (black line) and the host galaxy WISEA J073544.83+663717.3 (gray line). Vertical lines mark spectral features common in TDEs and AGNs, with hydrogen lines in red, helium lines (\ion{He}{2} $\lambda$ 4686, \ion{He}{1} $\lambda$ 5876, and \ion{He}{1} $\lambda$ 6678) in blue, nitrogen lines (\ion{N}{3} $\lambda$ 4100, \ion{N}{3} $\lambda$ 4640, and [\ion{N}{2}] $\lambda$ 6583) in green, and oxygen lines [(\ion{O}{3}] $\lambda$ 4959 and [\ion{O}{3}] $\lambda$ 5007) in orange. Both spectra have been binned in $\sim 9$ \AA \ bins and the dichroic crossover region has been masked out for the SNIFS spectrum.}
 \label{fig:spectrum}
\end{figure*}

After concluding that ATLAS18mlw was likely a TDE, we obtained a spectrum of the host galaxy WISEA J073544.83+663717.3 with the Gemini Multi-Object Spectrograph \citep[GMOS; ][]{hook04} on the 8.1-m Gemini North telescope. This spectrum was taken with the B600 grating and dithered both along the spatial and spectral directions to mitigate the presence of chip gaps and avoid the influence of bad pixels. The spectra were reduced using normal methods such as bias, flat-fielding, and wavelength calibration using arc lamps. The individual dithered spectra were median combined to minimize the effect of cosmic ray hits. Figure \ref{fig:spectrum} shows the SNIFS and GMOS spectra with common TDE and AGN emission lines marked. 

With this deeper and higher resolution GMOS spectrum we fit for the redshift the methods of \citet{hayden21}. This method consists of a weighted cross-correlation \citep{kelson00} between our observed spectrum and galaxy templates from SDSS\footnote{\url{http://classic.sdss.org/dr5/algorithms/spectemplates/}}. Using the weighted average of six binning choices from 5 to 20 \AA\, we find an average observed redshift of $z = 0.0734 \pm 0.0004$, which we will adopt for the remainder of our analysis. Stellar absorption lines such as the G-band (4304 \AA), \ion{Mg}{1} (5175 $\AA$), and \ion{Na}{1} (5890 $\AA$) were used to determine the redshift. This redshift implies a luminosity distance of 334 Mpc and an angular scale of $\sim 1.41$ kpc arcsec$^{-1}$.

\section{Host-Galaxy Properties}
\label{sec:host}

\begin{figure*}
\centering
 \includegraphics[width=.48\textwidth]{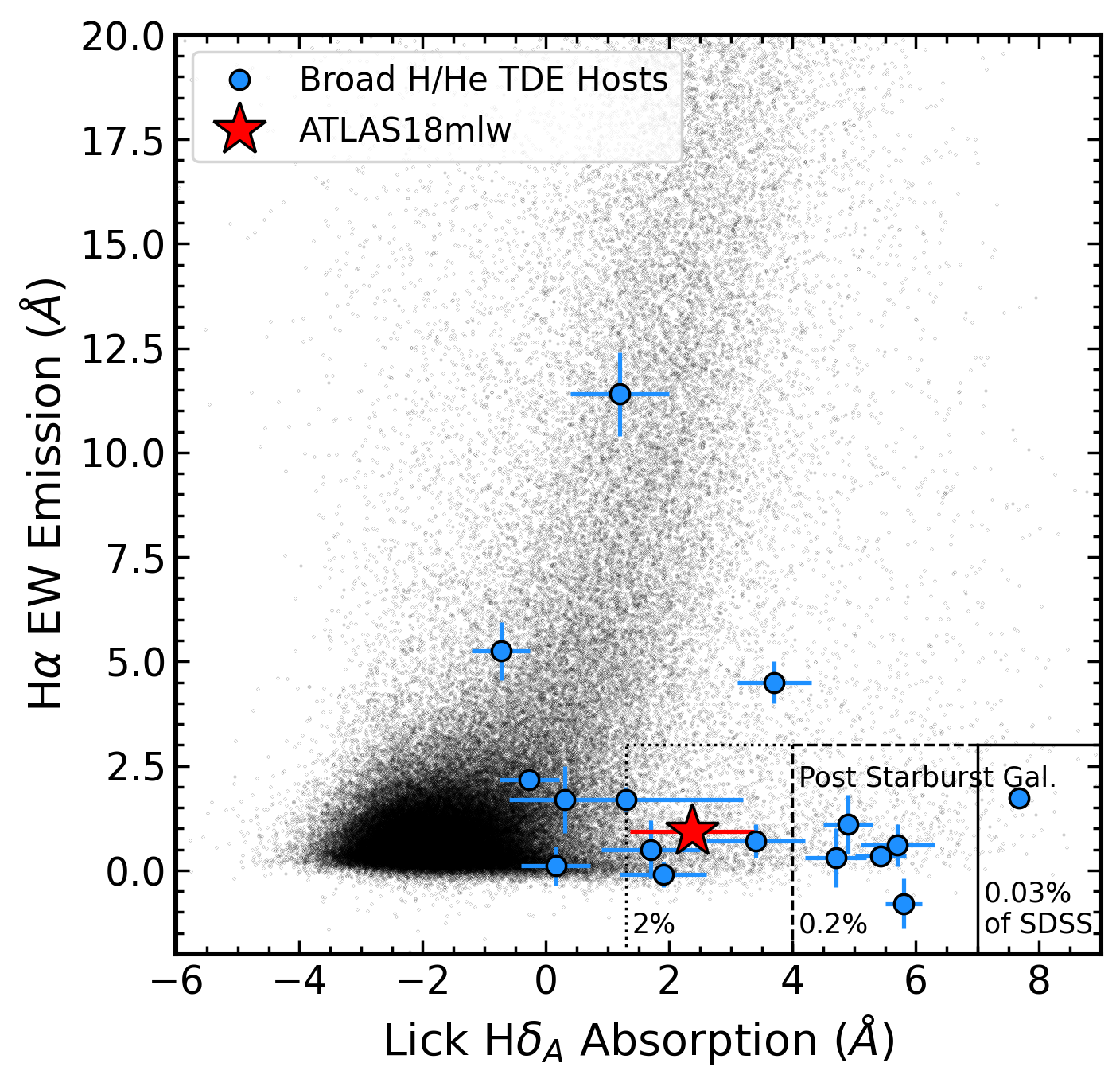}\hfill
 \includegraphics[width=.48\textwidth]{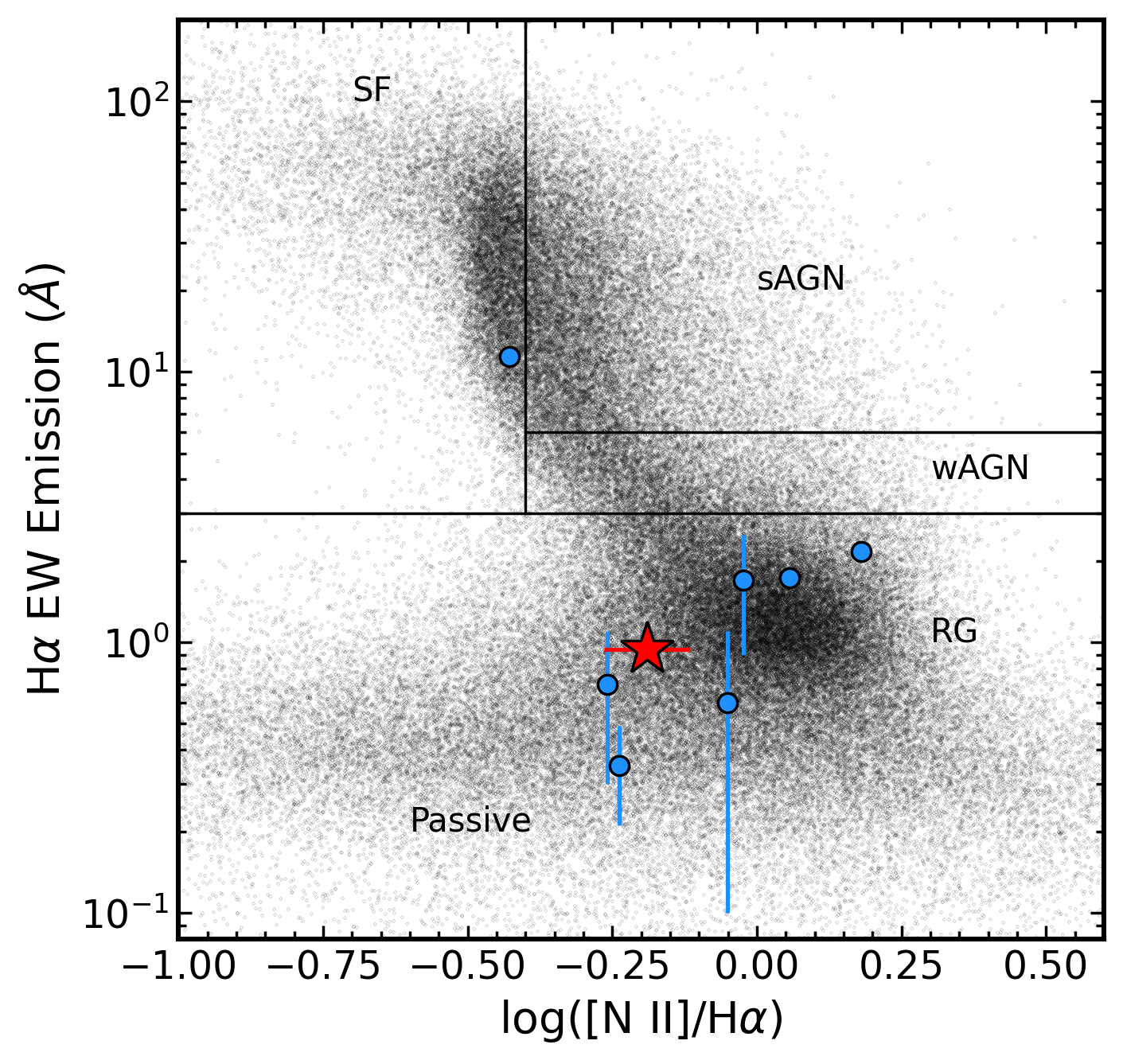} \\
 \includegraphics[width=.48\textwidth]{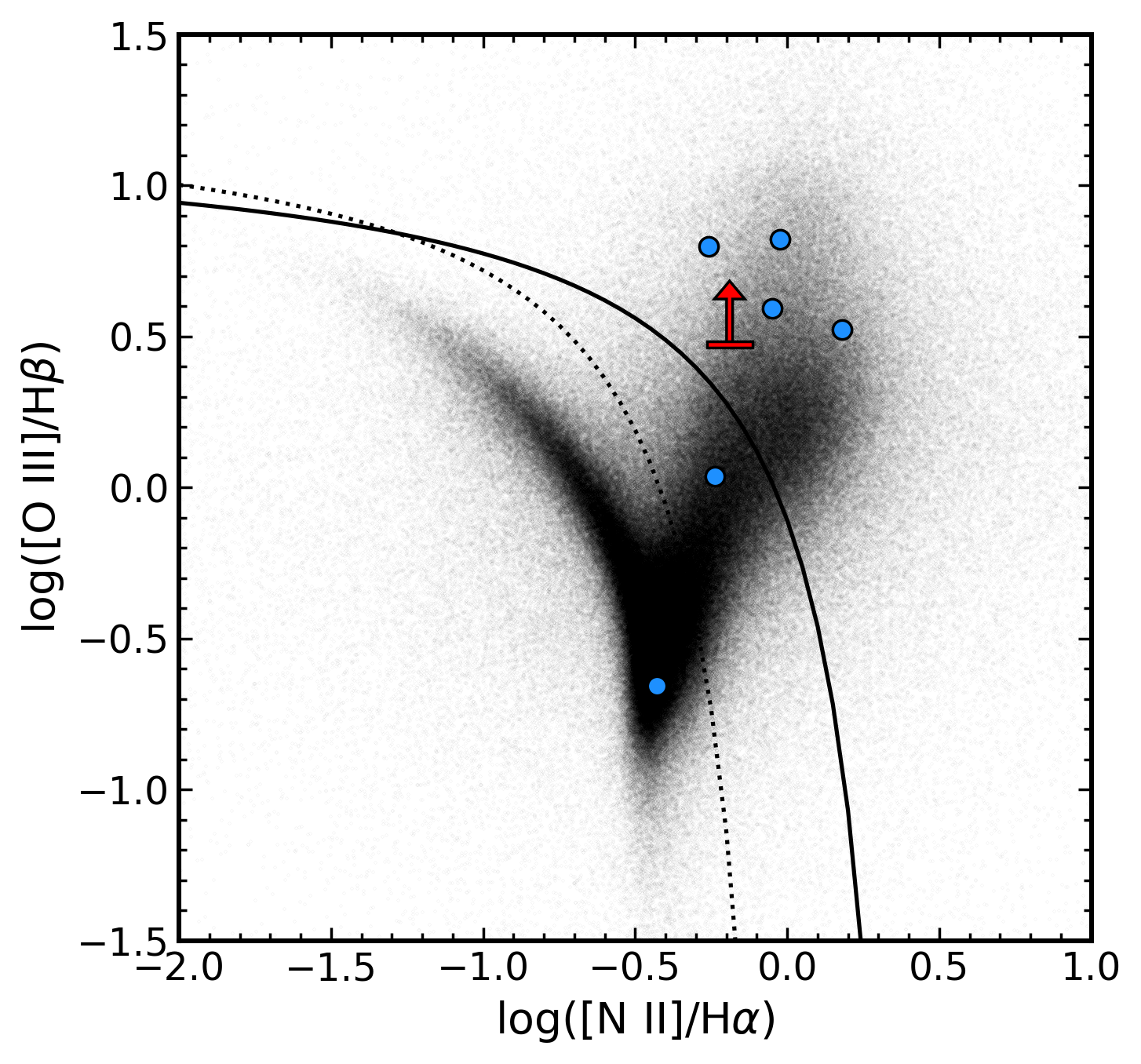}\hfill
 \includegraphics[width=.48\textwidth]{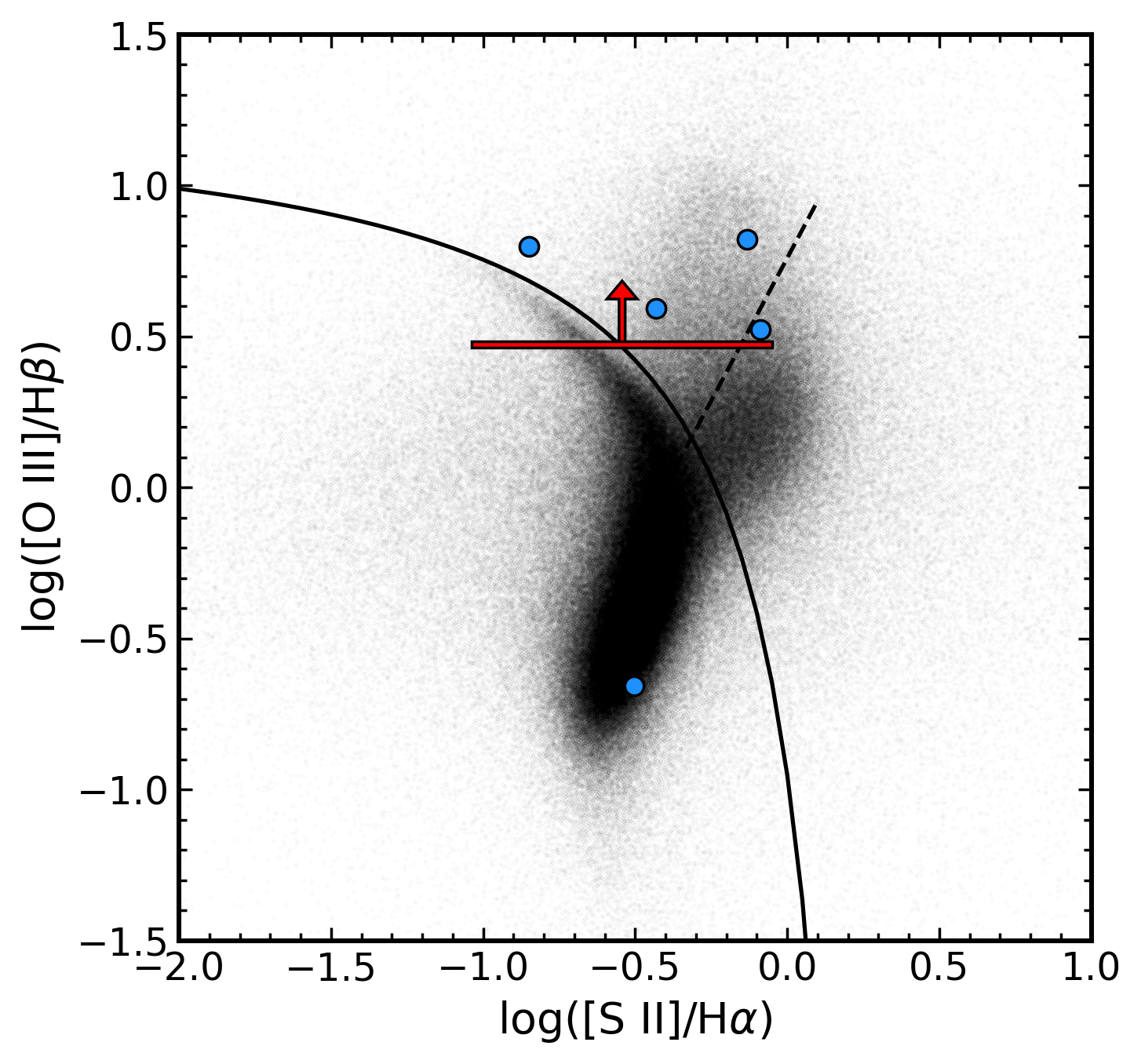}
 \caption{Line measurements of the host galaxy WISEA J073544.83+663717.3 computed from the GMOS spectrum. \textit{Upper Left Panel}: H$\alpha$ emission equivalent width (EW), tracing ongoing star formation, as compared to the Lick H$\delta_A$ absorption index, tracing star formation over the past Gyr. The host WISEA J073544.83+663717.3 is represented by a red star and other TDE hosts are shown with blue circles. WISEA J073544.83+663717.3 is a quiescent Balmer strong galaxy, similar to many TDE hosts. \textit{Upper Right Panel}: H$\alpha$ emission EW (W$_{H\alpha}$), as compared to log$_{10}$([\ion{N}{2}] / H$\alpha$), otherwise known as the WHAN diagram \citep{cidfernandes11}. Several lines separating star-forming galaxies (SF), strong AGN (sAGN), weak AGN (wAGN), and passive and ``retired galaxies'' (RG) are shown \citep{cidfernandes11}. \textit{Lower Left Panel}: log$_{10}$([\ion{O}{3}] / H$\beta$) vs. log$_{10}$([\ion{N}{2}] / H$\alpha$) diagram \citep{baldwin81, veilleux87}. The solid line theoretically separates AGNs (above) and H II-regions (below) \citep{kewley01}. The dotted line empirically separates the same classifications \citet{kauffmann03}. \textit{Lower Right Panel}: log$_{10}$([\ion{O}{3}] / H$\beta$) vs. log$_{10}$([\ion{S}{2}] / H$\alpha$) diagram \citep{veilleux87}. The solid line represents the same separation as before \citep{kewley01}. The diagonal dashed line separates Seyferts (above) and LINERs (below) \citep{kewley06}. WISEA J073544.83+663717.3 has only upper limits on the H$\beta$ emission. Galaxies from the SDSS Data Release 8 \citep{eisenstein11} are shown in black to illustrate the broader distribution of galaxy properties in these parameter spaces.}
 \label{fig:ew_bpt}
\end{figure*}

To better understand the host galaxy, we fit stellar population synthesis models to the archival photometry of WISEA J073544.83+663717.3 (shown in Table~\ref{tab:arch_phot}), excluding the $FUV$ limit. We made use of the Fitting and Assessment of Synthetic Templates \citep[\textsc{Fast};][]{kriek09} code to obtain an SED for the host galaxy. In this fit we assumed a \citet{cardelli89} extinction law with $\text{R}_{\text{V}} = 3.1$ and a foreground Galactic extinction of $\text{A}_{\text{V}} = 0.113$ mag \citep{schlafly11}, a Salpeter IMF \citep{salpeter55}, an exponentially declining star-formation rate, and the \citet{bruzual03} stellar population models. From our \textsc{Fast} fit, we find that WISEA J073544.83+663717.3 has a stellar mass of M$_* = 2.8^{+0.6}_{-0.4}$ $\times 10^9$ M$_{\odot}$, an age of $2.8^{+0.7}_{-0.6}$ Gyr, and a star formation rate of SFR $= 1.5^{+1.3}_{-1.1} \times 10^{-2}$ M$_{\odot}$ yr$^{-1}$.

From the commonly-used scaling relationship between stellar mass and SMBH mass \citep{mcconnell13, mendel14}, we estimate a central black hole mass of $\sim 10^{6.4}$ M$_{\odot}$, where these scaling relations have a typical scatter of $\sim 0.4$ dex. If we instead use the scaling of \citet{reines15}, we find a central black hole mass of $\sim 10^{5.8}$ M$_{\odot}$. As the scaling of \citet{reines15} includes more low-mass galaxies, we adopt this SMBH mass throughout the remainder of the manuscript. This mass estimate is close to the median BH mass on which TDEs occur \citep[$\sim 10^6 M_\odot$; ][]{wevers17, ryu20c} and is also consistent with the range of SMBH masses for which TDEs occur most efficiently \citep[$\leq 10^7 M_\odot$; ][]{vanvelzen18}.

The GMOS spectrum provides additional insight into the host-galaxy properties. We used the Penalized Pixel-Fitting code \citep[pPXF;][]{cappellari04, cappellari17} to simultaneously fit a stellar and gas component to the host galaxy spectrum, which allows for robust measurements of emission line fluxes. We measured the luminosity of the narrow H$\alpha$ line in the GMOS spectrum and applied the scaling between SFR and H$\alpha$ luminosity from \citet{osterbrock06} to obtain a SFR of $3.5 \times 10^{-3}$ M$_{\odot}$ yr$^{-1}$, which is smaller than, but consistent with, the value obtained from the SED fits. In addition to the bulk host galaxy properties, we can constrain the presence of AGN activity from the spectrum. From pPXF we measure an [\ion{O}{3}] 5007$\lambda$ emission line luminosity of $3.9 \times 10^{39}$~erg~s$^{-1}$. This is well below the median AGN [\ion{O}{3}] line luminosity indicating that the host galaxy does not harbor a strong AGN, but not ruling out the presence of low-luminosity AGN \citep[e.g.,][]{zakamska03, bongiorno10}. If the host galaxy harbors an AGN, we can use the bolometric luminosity scaling of \citet{pennell17} to estimate a bolometric luminosity of $\sim 3\times10^{44}$~erg~s$^{-1}$, similarly lying well below the median observed value \citep{woo02, lusso12}.

In addition to straightforward line-luminosity measurements, we make use of well-accepted emission-line diagnostic diagrams to understand the host galaxy properties and activity levels. Again using pPXF, we estimated emission line fluxes and uncertainties from the GMOS spectrum. In the upper left panel of Figure \ref{fig:ew_bpt} we show the  H$\alpha$ emission equivalent width (EW) as compared to the Lick H$\delta_A$ absorption index. The H$\alpha$ EW traces current star formation whereas the Lick H$\delta_A$ index traces star formation within the past Gyr. This diagram is commonly used to select post-merger and post-starburst systems based on their low current star formation rate but high recent star formation \citep[e.g.,][]{french18}. The host galaxy of ATLAS18mlw has Lick H$\delta_A$ $\simeq 2.4 \pm 1.0$ \AA, as measured by PyPHOT\footnote{\url{https://mfouesneau.github.io/docs/pyphot/\#}}, consistent with a quiescent Balmer strong galaxy, similar to many other TDE host galaxies.

In the upper right panel of Figure \ref{fig:ew_bpt} we show the H$\alpha$ emission EW (W$_{H\alpha}$), as compared to log$_{10}$([\ion{N}{2}] / H$\alpha$), also known as the WHAN diagram \citep{cidfernandes11}. This diagram separates actively star forming and active galaxies from passive galaxies using the H$\alpha$ emission EW. The galaxies with high W$_{H\alpha}$ are further separated by log$_{10}$([\ion{N}{2}] / H$\alpha$) into pure star-forming galaxies and active galaxies. The host galaxy of ATLAS18mlw resides in between the retired and passive galaxies regimes, suggesting that the host does not harbor a strong AGN. Again, the host galaxy is similar in terms of its location in the WHAN diagram to other TDE hosts.

In the bottom two panels of Figure \ref{fig:ew_bpt} we show the traditional log$_{10}$([\ion{N}{2}] $\lambda 6583$ / H$\alpha$) and log$_{10}$([\ion{S}{2}] $\lambda\lambda\ 6717, 6731$ / H$\alpha$) emission line ratio diagrams \citep{baldwin81, veilleux87}. H$\beta$ emission was not detected in the host galaxy spectrum, and we therefore use 3$\sigma$ upper limits. In both diagrams, the lower limit on log$_{10}$([\ion{O}{3}] $\lambda 5007$ / H$\beta$) lies in the AGN/Seyfert region, but similar to other TDE hosts \citep{graur18}. Addition, the large uncertainty on [\ion{S}{2}] flux leads to ambiguity on the classification of the source between a star-forming galaxy, a Seyfert, and a LINER \citep[low-ionization nuclear emission-line region;][]{heckman80}. Finally, if we assume a typical Balmer decrement of 2.86 \citep{osterbrock89} to estimate the H$\beta$ flux from the observed H$\alpha$ flux, we obtain a line ratio of log$_{10}$([\ion{O}{3}] / H$\beta$) $\simeq 0.9$.

We can also constrain the possibility of the host galaxy harboring a strong or obscured AGN through mid-infrared colors. Using the WISE photometry of the host galaxy, we find a W1 - W2 color of $0.22 \pm 0.17$ Vega mag. This is bluer than the 0.8 mag cut commonly imposed for AGNs \citep{assef13} and consistent with the host galaxies of many TDEs and ANTs \citep{hinkle21b}. While the WISE data cannot rule out an AGN on its own, it is unlikely that the galaxy hosts a strong or highly-obscured AGN.

Finally, we measured an X-ray flux at the position of the host galaxy using archival ROSAT \citep[ROentgen SATellite; ][]{voges99} images. We find a 3$\sigma$ upper limit of $< 0.03$ counts s$^{-1}$ in the $0.3 - 2.0$ keV bandpass. Assuming an AGN-like photon index of 1.75 \citep{ricci17} and the Galactic column density along the line of sight of $3.63 \times 10^{20}$ cm$^{-2}$ \citep{HI4PI16}, we find a $0.3 - 10$ keV limit of $<1.0 \times 10^{-12}$ erg  s$^{-1}$ cm$^{-2}$, which corresponds to a luminosity of $<1.3 \times 10^{43}$ erg  s$^{-1}$. This is fully consistent with a normal AGN, and cannot solely rule out the presence of a strong AGN. Unfortunately, besides ROSAT, there is little X-ray coverage of this host galaxy to further constrain the prescence of an AGN. Nevertheless, combining emission line diagnostics, MIR colors, and the X-ray upper limit suggest that WISEA J073544.83+663717.3 is unlikely to host a strong AGN.

\section{Analysis}
\label{sec:analysis}

While there is relatively little data during the transient event, we can nonetheless place constraints on the transient evolution and understand it in the context of other TDEs. Fitting the H$\alpha$ emission as a single Gaussian, we find a flux of $(7.3 \pm 1.1) \times 10^{-15}$ erg cm$^{-2}$ s$^{-1}$ corresponding to a line luminosity of $(9.6 \pm 1.4) \times 10^{40}$ erg s$^{-1}$. The equivalent width of the line relative to the nearby continuum is $112 \pm 17$ \AA. The width of the spectral line is  $(1.5 \pm 0.3) \times 10^{4}$ km s$^{-1}$, all similar to known TDEs \citep[e.g.,][]{vanvelzen20b, gezari21}. Finally, because of the location of the SNIFS dichroic crossover region at the redshift of ATLAS18mlw, it is not possibly to cleanly detect and measure the properties of the tentative H$\beta$ emission line.

A common analysis of optically-selected TDEs is blackbody fits to their UV/optical SEDs to obtain effective temperatures and bolometric luminosities. Unfortunately we do not have any UV photometry during the flare to probe the high-energy emission from this event directly. It is known that TDEs without UV photometry near their peaks can have sizable uncertainties on their SED properties \citep[e.g.,][]{hinkle20a}.

We can still make a rough estimate of these properties by fitting the optical spectrum as a blackbody. To do this, we first corrected the SNIFS spectrum for Galactic extinction. While host galaxy extinction is also important, it is difficult to estimate. Nevertheless, the WISE colors and lack of an observed \ion{Na}{1} D absorption line in both the SNIFS and GMOS spectra indicate that nuclear dust extinction is small \citep[e.g.,][]{poznanski12}. The GMOS host-galaxy spectrum does not cover a wide enough range in wavelength to host-subtract the SNIFS transient spectrum. We thus fit the spectrum below $\sim 4900$ \AA\ where the host galaxy is faint and a large majority of the flux is from the transient. Additionally, fitting below 4900 \AA\ allowed us to avoid the affects of the crossover region, where the SNIFS flux calibration is often poor and artifacts can be present.

\begin{figure*}
\centering
 \includegraphics[width=0.9\textwidth]{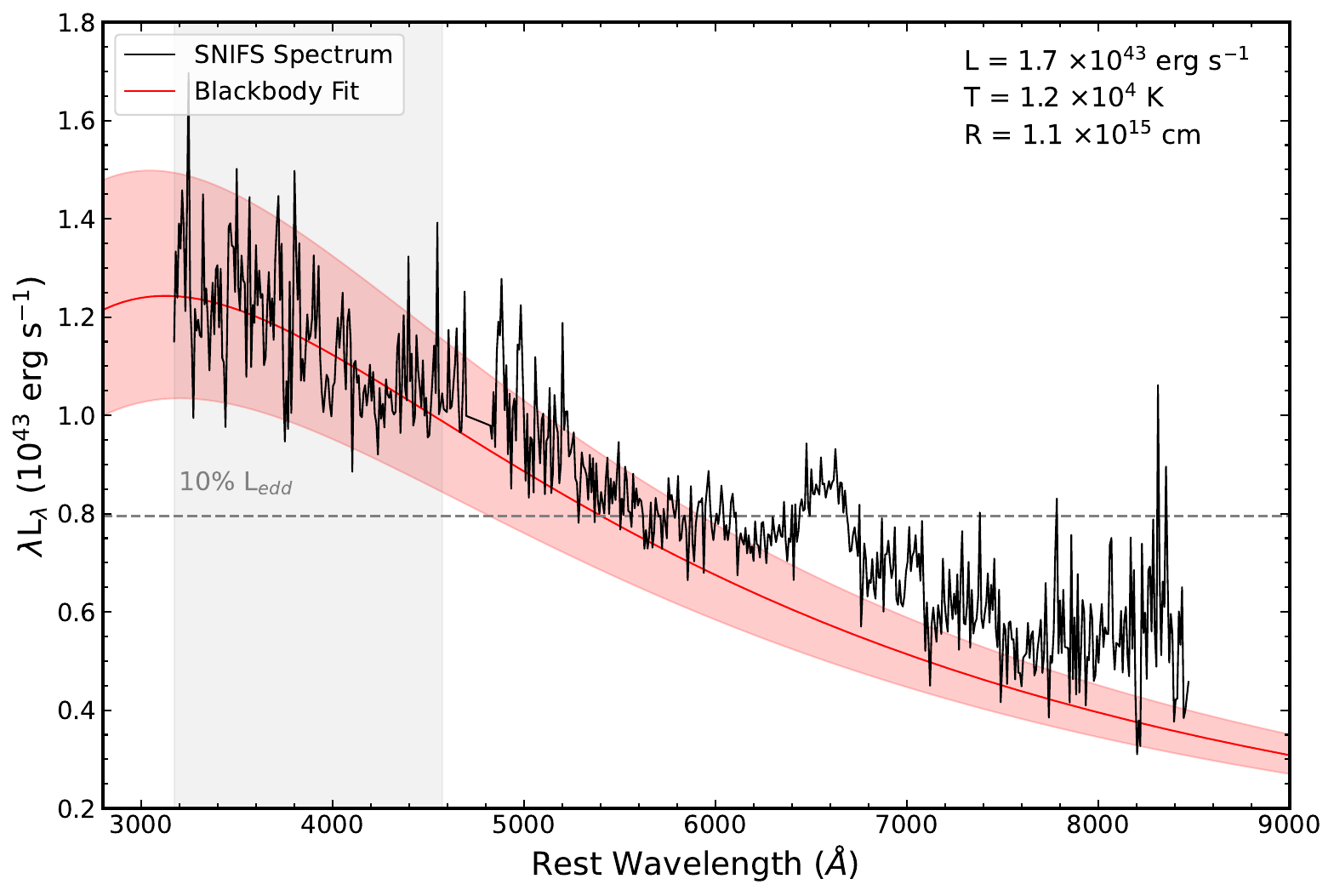}\hfill
 \caption{MCMC blackbody fit to the SNIFS spectrum of ATLAS18mlw. We present the spectrum in $\lambda L_{\lambda}$ to show that the TDE is clearly dominated by continuum emission in the blue and UV wavelengths. The best-fitting luminosity, temperature, and radius are shown in the upper right corner.  The gray shaded area represents the portion of the spectrum that has been fit as a blackbody. The gray dashed line represents 10 percent of the Eddington luminosity for a 10$^{5.8}$ M$_\odot$ SMBH. As before, the spectrum has been binned in $\sim 9$ \AA \ bins and the dichroic crossover region has been masked out.} 
 \label{fig:BB_fit}
\end{figure*}

We fit the TDE spectrum with a Markov Chain Monte Carlo (MCMC) approach with flat temperature priors of 1000 K $<$ T $<$ 55000 K. We assumed a blackbody model for our fit and obtained the bolometric luminosity, radius, and effective temperature. The $\chi^2$ per degree of freedom for this fit is 1.0, indicating a good fit to the blue portion of the spectrum. The spectrum and corresponding blackbody fit are shown in Figure \ref{fig:BB_fit}. The fitted luminosity of L $= (1.7 \pm 0.1) \times 10^{43}$ erg s$^{-1}$, radius of R $= (1.1 \pm 0.1) \times 10^{15}$ cm, and temperature of T $= (1.2  \pm 0.1) \times 10^{4}$ K are also shown on the figure. These uncertainties are very likely underestimated as they only represent the statistical uncertainty on the fact and ignore the systematic affects associated with fitting only optical data for hot sources. As expected from the lack of host subtraction, the spectrum deviates from the blackbody fit at redder wavelengths.

Using the blackbody parameters computed above, we can better understand the light curve of ATLAS18mlw in the context of other TDEs. Similar to previous TDEs \citep{holoien20, hinkle20a, hinkle21b}, we bolometrically corrected the optical light curves shown in Fig. \ref{fig:opt_lc} using the fitted luminosity at the time of the SNIFS spectrum. This was done by scaling the ATLAS $c$ and ZTF $g$ light curves to match the bolometric luminosity estimated from the SNIFS spectrum. As there was a single bolometric luminosity estimate, the scaling between the $cg$ data and the bolometric luminosity was assumed to be constant throughout the TDE evolution. This assumes a relatively flat temperature evolution, which is often acceptable for a TDE \citep{hinkle21a, vanvelzen21, gezari21}. 

\begin{figure*}
\centering
 \includegraphics[width=0.85\textwidth]{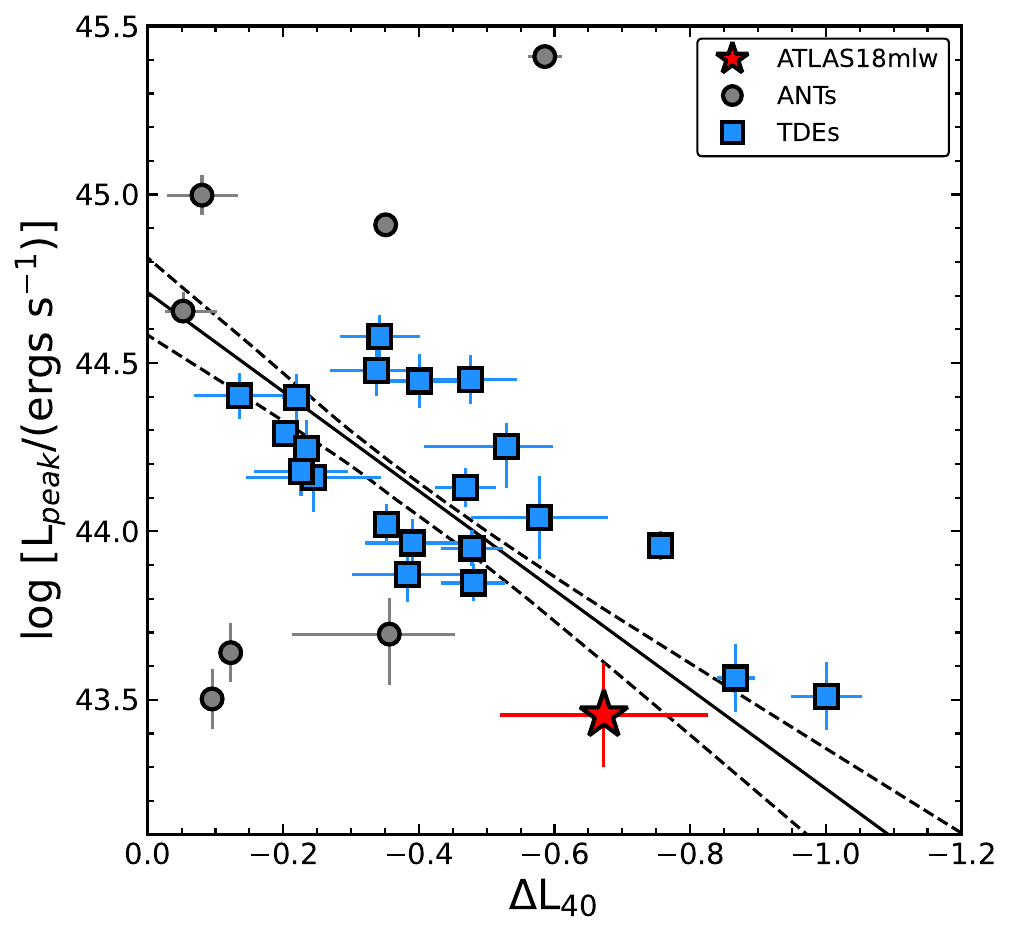}\hfill
 \caption{Peak bolometric luminosity versus the decline rate of several TDEs (blue squares) and ANTs (gray circles). The decline rate $\Delta$L$_{40}$ is the difference between the log of the peak luminosity and the log of the luminosity at 40 days after peak \citep{hinkle20a}. ATLAS18mlw is shown with a red star, and the uncertainties here are 90\% confidence intervals.}
 \label{fig:deltaL40}
\end{figure*}

Following the procedure outlined in \citet{hinkle20a}, we computed the peak luminosity and decline rate measured at 40 days ($\Delta L_{40}$) for ATLAS18mlw. Given the small number of epochs near peak light, we used a Monte Carlo approach to estimate the peak luminosity, fitting the data near peak 5000 times, each time with the luminosities perturbed by their uncertainties assuming Gaussian errors. This yielded a peak luminosity of log(L [erg s$^{-1}$]) = $43.5 \pm 0.2$. To ensure a robust uncertainty estimate given the lack of UV data, this error represents the 90\% confidence interval of the luminosities computed from our Monte Carlo estimation. Estimating a time of peak light with the same procedure yields MJD$_{peak} = 58204.4 \pm 6.8$, consistent with the approximate optical peak seen in Fig. \ref{fig:opt_lc}. Given the SMBH mass estimated from the host galaxy SED fitting, this represents an Eddington ratio of $\sim 20$\%, similar to other TDEs \citep[e.g.,][]{wevers17, mockler19}. We next computed the decline rate and found $\Delta L_{40} = -0.7 \pm 0.2$. The peak luminosity and decline rate of ATLAS18mlw, along with well-studied TDEs and ambiguous nuclear transients (ANTs) are shown in Figure \ref{fig:deltaL40}, with ATLAS18mlw broadly following the trend seen for TDEs \citep[e.g.,][]{hinkle20a, hinkle21b}.

\section{Discussion} \label{sec:discusssion}

The observational data of ATLAS18mlw appears similar to many other TDEs, with no significant archival host variability, a likely post-merger host galaxy, a strong blue continuum with broad emission lines near peak, and a timescale of a few months in the optical. While we do not have direct measurements of the UV flux, the presence of strong H$\alpha$ emission indicates that there must be significant far UV flux in the neighborhood of the transient, typical of TDEs \citep[e.g.,][]{roth18} and Seyfert galaxies \citep[e.g.,][]{antonucci93}. Transients associated with AGN accretion rate changes also commonly exhibit broad H$\alpha$ emission \citep[e.g.,][]{gezari17a, neustadt20, frederick20}. As such, we must consider the possibility that ATLAS18mlw is a smooth AGN flare. Firstly, the short overall timescale of ATLAS18mlw is less than the roughly year-long timescales of AGN-related flares \citep{hinkle21b}. Additionally, the spectrum of ATLAS18mlw is well-fit as blackbody, consistent with TDEs \citep[e.g.,][]{holoien14b, holoien16a}, whereas most AGN are better fit by power-laws \citep[e.g.,][]{vandenberk01, neustadt20}. Finally, the lack of significant optical variability, weak narrow H$\alpha$ emission from the host galaxy, and the relatively blue WISE colors are inconsistent with the presence of an AGN.

ATLAS18mlw also shares some characteristics with supernovae, that can occur in the nuclei of their host galaxies. Most similar to ATLAS18mlw in terms of absolute magnitude are Type Ia supernovae \citep[e.g.,][]{folatelli10}. However, the optical spectrum of ATLAS18mlw does not have any of the expected absorption lines seen for any class of Type I supernova \citep{filippenko97}. Conversely, Type II supernovae do show an early blue continuum and broad H$\alpha$ emission. However, the photometric evolution of ATLAS18mlw is unlike typical Type II supernova, being both more luminous and faster-evolving than the large majority of Type II supernovae \citep[e.g.,][]{galbany16}. Additionally, we note that the light curves of supernovae occurring in an AGN disk may be similar to the light curve of ATLAS18mlw in terms of timescale and luminosity \citep{grishin21}. However, theoretical predictions on the spectra of these transients are needed to disentangle such a scenario from the TDE scenario favored here due to the agreement between the properties of ATLAS18mlw and its host galaxy with known TDEs.

With ATLAS18mlw as a TDE candidate rather than an AGN flare or supernova, we can compare its properties to well-observed TDEs in the literature. Using the optical spectroscopic classification scheme of \citet{leloudas19} and \citet{vanvelzen21}, this TDE would likely be a TDE-H due to the dominance of H$\alpha$ over any other spectral feature. TDE-H objects tend to have broader lines than TDE-Bowen objects \citep{charalampopoulos21}, consistent with the H$\alpha$ line width computed for ATLAS18mlw. The $\sim 15,000$ km s$^{-1}$ H$\alpha$ line seen for ATLAS18mlw lies at the median value seen for TDE-H objects, and a factor of a few higher than the median H$\alpha$ line width for TDE-Bowen objects \citep{charalampopoulos21}. However, some TDEs, like ASASSN-15oi \citep{holoien18a} and ASASSN-19dj \citep{hinkle21a} show signs for changes in spectral features and the properties of observed emission lines, which we cannot probe for ATLAS18mlw with our single spectrum. Nevertheless, the appearance of the broad H$\alpha$ line at an approximate phase of nine days before peak for ATLAS18mlw is among the earliest detected. Many TDEs show an increase in line luminosity near or shortly after peak light before declining \citep{charalampopoulos21}, although here we only have a single spectrum with which to constrain the line emission.

The temperature derived from our blackbody fit, log(T[K]) = $4.06 \pm 0.01$, is towards the low end of temperatures for the TDE-H subclass \citep{vanvelzen21, hammerstein22}. However, the lack of UV photometry makes the systematic errors on our temperature significantly higher than the statistical errors from the blackbody fit alone. Nonetheless, both the TDE-He and TDE-Bowen classes have much higher median temperatures, by $\sim 35$\% and $\sim 40$\% respectively, thus supporting this object belonging to the TDE-H class. Additionally, the estimated blackbody radius of log(R [cm]) = $15.07 \pm 0.02$ is fully consistent with the radii of the TDE-H objects. \citet{vanvelzen21} find that the separation between TDE-H and TDE-Bowen is strongest for the blackbody radius, again supporting a TDE-H nature. Still, if our crude blackbody fit without UV photometry underestimates the temperature, then the blackbody radius may be overestimated for a similar luminosity.

In terms of the bolometric evolution, we find that ATLAS18mlw is most similar to the ``faint and fast'' TDEs iPTF16fnl \citep{blagorodnova17, brown18}, ZTF19abzrhgq \citep[AT2019qiz; ][]{nicholl20}, AT2020neh \citep{angus22}, and AT2020wey \citep{charalampopoulos22}. Of these TDEs, ATLAS18mlw has the second lowest peak luminosity, although the bolometric luminosity estimate is the weakest as there is no UV data to confirm the temperature. ATLAS18mlw is also the only likely TDE-H object of these faint and fast TDEs. The decline rate of ATLAS18mlw indicates that it is a fast evolving TDE, with only a small number of sources declining faster. These are iPTF16fnl, ZTF19abzrhgq, ASASSN-14ae \citep{holoien14b}, and AT2020wey. Several studies have suggested that faint and fast TDEs may be tied to lower mass SMBHs \citep{blagorodnova17}, potentially as a result of more efficient outflow launching around such SMBHs \citep{nicholl20}. The SMBH mass for ATLAS18mlw is similar to, but slightly lower than other TDEs \citep{wevers17, nicholl22}, although the estimate is only based on the total stellar mass from SPS fitting, and therefore has a large uncertainty.

It is also interesting to note that the position of ATLAS18mlw in peak-luminosity/decline-rate space is much more similar to the TDEs than any of the ambiguous nuclear transients (ANTs) shown on Fig. \ref{fig:deltaL40}. Recently, \citet{charalampopoulos22} measured the decline rate of another faint and fast TDE AT2020wey and find it to be $\Delta$L$_{40}$ = $-1.28$, fully consistent with this relationship. These results support using the log(L$_{peak}$) -- $\Delta$L$_{40}$ parameter space to help us differentiate between different classes of transients even without abundant data during the event, as will be the case for many discoveries by the Legacy Survey of Space and Time \citep[LSST; ][]{ivezic19} on the Vera Rubin Observatory.

For the case of ATLAS18mlw, the late identification as a TDE precluded any follow-up observations, highlighting the need for rapid response to newly-discovered nuclear transients. As the discovery of TDEs continues to expand, it is important to prioritize data that best helps us understand physics driving the observed emission.  Optical spectroscopy and high-energy UV/X-ray photometry \citep[e.g.,][]{mushotzky18, kulkarni21} ideally beginning prior to peak light, are necessary to understand the detailed properties of individual events. 

This goal requires the expansion of spectroscopic classification surveys to match the growing flood of transient discoveries each night. Even now, the number of discovered transients is roughly an order of magnitude higher than the number of transients classified spectroscopically. One obvious improvement is the robotizing of existing classification surveys to improve efficiency. For instance, the SCAT survey is currently classically scheduled and observed with 323 classifications in the last year, representing $\sim 13$\% of all classified transients. After the ongoing robotizing of the UH88 telescope\footnote{This effort is funded by an NSF Major Research Instrumentation Program (MRI) grant (Award Number: 1920392).} is completed in late 2022, this number is expected to increase to $\sim 1200$ per year.

Sources like ATLAS18mlw show that even with minimal data, important conclusions can be drawn and comparisons to very well-studied objects can be made. In particular, the host galaxies of sources like ATLAS18mlw can and should be incorporated into TDE population studies as these samples grow, although quantification of the reliability of source classifications as TDEs will be important.

\acknowledgments

We thank the referee for comments that have improved the paper. We also thank Alexa Anderson for helpful comments on the manuscript.

J.T.H. and this work was supported by NASA award 80NSSC21K0136. M.A.T. acknowledges support from the DOE CSGF through grant DE-SC0019323. Support for T.W.-S.H. was provided by NASA through the NASA Hubble Fellowship grant HST-HF2-51458.001-A awarded by the Space Telescope Science Institute, which is operated by the Association of Universities for Research in Astronomy, Inc., for NASA, under contract NAS5-26555.  B.J.S is supported by NSF grants AST-1908952, AST-1920392, AST-1911074, and NASA award 80NSSC19K1717. Support for G.A. was provided by the Director, Office of Science, Office of High Energy Physics of the U.S. Department of Energy under Contract No. DE-AC025CH11231

This work uses data from the University of Hawaii's ATLAS project, funded through NASA grants NN12AR55G, 80NSSC18K0284, and 80NSSC18K1575, with contributions from the Queen's University Belfast, STScI, the South African Astronomical Observatory, and the Millennium Institute of Astrophysics, Chile.

This paper includes data collected by the TESS mission, which are publicly available from the Mikulski Archive for Space Telescopes (MAST). Funding for the TESS mission is provided by NASA’s Science Mission directorate.

Parts of this research were supported by the Australian Research Council Centre of Excellence for All Sky Astrophysics in 3 Dimensions (ASTRO 3D), through project number CE170100013.

This work is based on observations made by ATLAS and Gemini. The authors wish to recognize and acknowledge the very significant cultural role and reverence that the summits of Haleakal\=a and Maunakea have always had within the indigenous Hawaiian community.  We are most fortunate to have the opportunity to conduct observations from these mountains.

\section*{Data availability}
	
The data underlying this article will be shared on reasonable request to the corresponding author.

\facilities{ASAS-SN \citep{shappee14, kochanek17}, ATLAS \citep{tonry18}, GMOS \citep{hook04}, SNIFS \citep{lantz04}, ZTF \citep{bellm19}}

\software{Linmix \citep{kelly07}, Matplotlib \citep{matplotlib}, Numpy \citep{numpy}, PYPHOT, pPXF \citep{cappellari04, cappellari17}, Scipy \citep{scipy}}

\bibliography{bibliography}{}
\bibliographystyle{aasjournal}

\end{document}